\documentstyle[12pt,epsf]{article}        

\renewcommand{\thefootnote}{\fnsymbol{footnote}}
 1

\catcode`\@=11 
\makeatletter
\def\@seccntformat#1{\csname the#1\endcsname.\hskip 1em}

\makeatother

\catcode`@=11
\def\chkspace{%
  \relax   
  \begingroup\ifhmode\aftergroup\dochksp@ce\fi\endgroup}
\def\dochksp@ce{%
  \unskip              
  \futurelet\chkspct@k\d@chkspc  
}
\def\d@chkspc{%
  \let\nxtsp@ce=\relax
  \ifx\chkspct@k.\else     
    \ifx\chkspct@k,\else
      \ifx\chkspct@k;\else
        \ifx\chkspct@k!\else
          \ifx\chkspct@k?\else
            \ifx\chkspct@k:\else
              \ifx\chkspct@k)\else
              \ifx\chkspct@k(\else
                \ifx\chkspct@k]\else
                  \ifx\chkspct@k-\else
                    \ifx\chkspct@k\egroup\else  
                      \let\nxtsp@ce=\put@space  
                    \fi
                  \fi
                \fi
              \fi
              \fi
            \fi
          \fi
        \fi
      \fi
    \fi
  \fi
  \nxtsp@ce
}
\def\put@space{$\;$}
\catcode`@=12

\def\ra{{$\rightarrow$}\chkspace}
\def\etal{{\it et al.}\chkspace}

\def\ie{{\it i.e.}\chkspace}

\def\eg{{\it eg.}\chkspace}

\def\apriori{{\it a priori}\chkspace}

\def\ep{{e$^+$e$^-$}\chkspace}

\def\gluino{\relax\ifmmode \tilde{g} \else $\tilde{g}$ \fi\chkspace}

\def\qq{\relax\ifmmode q\overline{q}
\else $q\overline{q}$ \fi\chkspace}

\def\QQ{$Q\overline{Q}$\chkspace}

\def\bb{\relax\ifmmode b\bar{b}
       \else $b\bar{b}$ \fi\chkspace}
\def\ccrm{\relax\ifmmode {\rm c}\bar{\rm c}
       \else ${\rm c}\bar{\rm c}$ \fi\chkspace}

\def\tt{\relax\ifmmode {\rm t}\bar{\rm t}
       \else ${\rm t}\bar{\rm t}$ \fi\chkspace}
\def\ss{\relax\ifmmode {\rm s}\bar{\rm s}
       \else ${\rm s}\bar{\rm s}$ \fi\chkspace}
\def\uu{\relax\ifmmode {\rm u}\bar{\rm u}
       \else ${\rm u}\bar{\rm u}$ \fi\chkspace}
\def\dd{\relax\ifmmode {\rm d}\bar{\rm d}
       \else ${\rm d}\bar{\rm d}$ \fi\chkspace}

\def\qqg{\relax\ifmmode q\overline{q}g
\else $q\overline{q}g$ \fi\chkspace}
\def\bbg{\relax\ifmmode b\overline{b}g
\else $b\overline{b}g$ \fi\chkspace}
\def\ccg{$c\overline{c}g$\chkspace}

\def\QQg{$Q\overline{Q}g$\chkspace}

\def\afb{\relax\ifmmode A_{FB} \else
{{$A_{FB}$}}\fi\chkspace}
\def\afbb{\relax\ifmmode A_{FB}^b \else
{{$A_{FB}^b$}}\fi\chkspace}
\def\pafb{\relax\ifmmode \tilde{A}_{FB} \else
{{$\tilde{A}_{FB}$}}\fi\chkspace}
\def\pafbb{\relax\ifmmode \tilde{A}_{FB}^b \else
{{$\tilde{A}_{FB}^b$}}\fi\chkspace}

\def\pafbzo{\relax\ifmmode \tilde{A}_{FB}|_{O(0)} \else
{{$\tilde{A}_{FB}|_{O(0)}$}}\fi\chkspace}
\def\pafbfo{\relax\ifmmode \tilde{A}_{FB}|_{\oalp} \else
{{$\tilde{A}_{FB}|_{\oalp}$}}\fi\chkspace}
\def\pafbso{\relax\ifmmode \tilde{A}_{FB}|_{\oalpsq} \else
{{$\tilde{A}_{FB}|_{\oalpsq}$}}\fi\chkspace}
\def\pafbto{\relax\ifmmode \tilde{A}_{FB}|_{\oalpc} \else
{{$\tilde{A}_{FB}|_{\oalpc}$}}\fi\chkspace}

\def\pafbbzo{\relax\ifmmode \tilde{A}_{FB}^b|_{O(0)} \else
{{$\tilde{A}_{FB}^b|_{O(0)}$}}\fi\chkspace}
\def\pafbbfo{\relax\ifmmode \tilde{A}_{FB}^b|_{\oalp} \else
{{$\tilde{A}_{FB}^b|_{\oalp}$}}\fi\chkspace}
\def\pafbbso{\relax\ifmmode \tilde{A}_{FB}^b|_{\oalpsq} \else
{{$\tilde{A}_{FB}^b|_{\oalpsq}$}}\fi\chkspace}
\def\pafbbto{\relax\ifmmode \tilde{A}_{FB}^b|_{\oalpc} \else
{{$\tilde{A}_{FB}^b|_{\oalpc}$}}\fi\chkspace}

\def\afbo0{\tilde{A}_{FB}|_{O(0)}}
\def\afbo1{\tilde{A}_{FB}|_{\oalp}}
\def\afbo2{\tilde{A}_{FB}|_{\oalpsq}}
\def\afbo3{\tilde{A}_{FB}|_{\oalpc}}

\def\lam{\relax\ifmmode \Lambda_{\overline{MS}}
       \else {{$\Lambda_{\overline{MS}}$}}\fi\chkspace}
\def\lamuds{\relax\ifmmode \Lambda^{(3)}_{\overline{MS}}
       \else {{$\Lambda^{(3)}_{\overline{MS}}$}}\fi\chkspace}
\def\lamudsc{\relax\ifmmode \Lambda^{(4)}_{\overline{MS}}
       \else $\Lambda^{(4)}_{\overline{MS}}$\fi\chkspace}
\def\lamudscb{\relax\ifmmode \Lambda^{(5)}_{\overline{MS}}
       \else $\Lambda^{(5)}_{\overline{MS}}$\fi\chkspace}

\def\alp{\relax\ifmmode \alpha_s\else $\alpha_s$\fi\chkspace}
\def\alpbar{\relax\ifmmode \bar{\alpha_s}
       \else $\bar{\alpha_s}$\fi\chkspace}
\def\alpmz{\relax\ifmmode \alpha_s(M_Z)\else
$\alpha_s(M_Z)$\fi\chkspace}
\def\alpmzsq{\relax\ifmmode \alpha_s(M_Z^2)
       \else $\alpha_s(M_Z^2)$\fi\chkspace}

\def\oalp{\relax\ifmmode O(\alpha_s)\else{{O($\alpha_s$)}}\fi\chkspace}
\def\oalpsq{\relax\ifmmode O(\alpha_s^2)
           \else{{O($\alpha_s^2$)}}\fi\chkspace}
\def\oalpc{\relax\ifmmode O(\alpha_s^3)
           \else{{O($\alpha_s^3$)}}\fi\chkspace}
\def\oalpf{\relax\ifmmode O(\alpha_s^4)
           \else{{O($\alpha_s^4$)}}\fi\chkspace}

\def\rb{\relax\ifmmode R_3^b/R_3^{all}
           \else{{$R_3^b/R_3^{all}$}}\fi\chkspace}
\def\rc{\relax\ifmmode R_3^c/R_3^{all}
           \else{{$R_3^c/R_3^{all}$}}\fi\chkspace}
\def\ruds{\relax\ifmmode R_3^{uds}/R_3^{all}
           \else{{$R_3^{uds}/R_3^{all}$}}\fi\chkspace}
\def\ri{\relax\ifmmode R_3^i/R_3^{all}
           \else{{$R_3^i/R_3^{all}$}}\fi\chkspace}
\def\rj{\relax\ifmmode R_3^j/R_3^{all}
           \else{{$R_3^j/R_3^{all}$}}\fi\chkspace}
\def\alpi{\relax\ifmmode \alpha^i_s/\alpha^{all}_s
           \else{{$\alpha^i_s/\alpha^{all}_s$}}\fi\chkspace}

\def\mbz{\relax\ifmmode m_b(M_Z)
           \else{{$m_b(M_Z)$}}\fi\chkspace}
\def\mbb{\relax\ifmmode m_b(M_b)
           \else{{$m_b(M_b)$}}\fi\chkspace}

\def\plb{Phys. Lett.\chkspace}
\def\npb{Nucl. Phys.\chkspace}

\def\prl{Phys. Rev. Lett.\chkspace}
\def\prd{Phys. Rev.\chkspace}
\def\zpc{Z. Phys.\chkspace}

\def\z0{{$Z^0$}\chkspace}
\def\Dst{\relax\ifmmode {\rm D}^* \else {D$^*$}\fi\chkspace}
\def\Dpl{\relax\ifmmode {\rm D}^+ \else {D$^+$}\fi\chkspace}
\def\D0{\relax\ifmmode {\rm D}^0 \else {D$^0$}\fi\chkspace}
\def\Kst{\relax\ifmmode {\rm K}^* \else {K$^*$}\fi\chkspace}
\def\K0{\relax\ifmmode {\rm K}^0_s \else {K$^0_s$}\fi\chkspace}
\def\Kpl{\relax\ifmmode {\rm K}^+ \else {K$^+$}\fi\chkspace}
\def\Kstz{\relax\ifmmode {\rm K}^{*0} \else {K$^{*0}$}\fi\chkspace}

\def\beq{\begin{equation}}
\def\eeq{\end{equation}}
\def\bea{\begin{eqnarray}}
\def\eea{\end{eqnarray}}
\newcommand{\eq}[1]{Eq.~(\ref{#1})}
\newcommand{\as}{\alpha_s}
\newcommand{\smin}{s_{\rm min}}
\newcommand{\msbar}{$\overline{\rm MS}$}
\begin{document}

\begin{flushright}
 SLAC--PUB--7915\\
 OUNP--99--06\\
Saclay/SPhT-T99/047\\
PITHA 99/12\\
May 1999\\
\end{flushright}

\vskip 1truecm
 
\begin{center}
{\large\bf
 MEASUREMENT OF THE RUNNING $b$-QUARK MASS \\
 USING e$^+$e$^-$ \ra \bbg  EVENTS
}

\vskip .8truecm

A. Brandenburg$^1$,
P.N. Burrows$^2$,
D. Muller$^3$,
N. Oishi$^4$,
P. Uwer$^5$

\end{center}
 
\vskip 1truecm
 
\begin{center}
{\bf ABSTRACT }
\end{center}
 
\noindent
We have studied the determination of the running $b$-quark mass, \mbz,
using \z0 decays into 3 or more hadronic jets.
We calculated the ratio of
$\geq3$-jet fractions in \ep \ra \bb vs. \ep \ra $q_l\overline{q_l}$ 
($q_l$ = $u$ or $d$ or $s$) events at next-to-leading order in
perturbative QCD using
six different infra-red- and collinear-safe jet-finding algorithms.
We compared with 
corresponding measurements from the SLD Collaboration and found a 
significant algorithm-dependence of the fitted \mbz value. 
Our best estimate, taking correlations into account, is
$m_b(M_Z)$ = $2.56 \pm 0.27$ {\rm (stat.)} $^{+0.28}_{-0.38}$ {\rm (syst.)}
$^{+0.49}_{-1.48}$  (theor.) GeV/$c^2$.


\vfil

{\footnotesize

\noindent
$^1$ Institut f\"ur Theoretische Physik, RWTH Aachen,
D-52056 Aachen, Germany and DESY Theory Group, D-22603 Hamburg, Germany.

\noindent
$^2$ Oxford University, Particle and Nuclear Physics, Oxford, OX1 3RH,
UK.

\noindent
$^3$ Stanford Linear Accelerator Center, 2575 Sand Hill Road,  Menlo
Park, CA 94025, USA.

\noindent
$^4$ Nagoya University, Nagoya 464, Japan.

\noindent
$^5$ Service de Physique Th\'eorique,
Centre d'Etudes de Saclay, F-91191 Gif-sur-Yvette cedex, France.

}

\eject
\setcounter{footnote}{0}
\renewcommand{\thefootnote}{\arabic{footnote}}
\section{Introduction}

\vskip .5truecm

\noindent
Three-jet events of the type \ep \ra $q{\bar q}g$ provide an ideal
laboratory for
making precise tests of Quantum Chromodynamics (QCD)~\cite{qcd}.
Since the initial state is free of strongly-interacting particles
the experimental environment is intrinsically `clean',
and the process is more amenable to calculation using perturbation
theory than, for example, multijet final states in hadron-hadron
or lepton-hadron collisions. A number of perturbative QCD (pQCD) 
predictions for 3-jet dominated
hadronic event-shape observables,
for massless quarks, complete at next-to-leading order (NLO)
are available~\cite{ert,nlo,GiGl92,CaSe96b,CaSe97,FrKuSi96,NaTr97}.

One would expect the emission of gluon radiation in events containing
massive quarks, \ep \ra \QQ$\;$
($Q$ = $b$ or $c$),
to be modified relative to that in \ep \ra 
$q_l\overline{q_l}$ ($q_l$ = $u$ or $d$ or $s$)
events due to the restriction in phase space imposed by the non-zero
quark mass. One would also expect such a modification to depend on
the choice of the event-shape observable, and to be potentially
relatively large for
those observables in which the quark-jet mass enters kinematically into
the definition.
Recently NLO calculations of \ep \ra \QQg have been performed
in which quark mass effects have been
taken into account explicitly~\cite{arnd,rodrigo,nason}.
>From these calculations one expects the size of the $c$-mass effects in 
\ccg events at $\sqrt{s}=M_Z$ to be rather small, at the level of 1\%.
However, for \bbg events the relative size of the $b$-mass effects on
event-shape observables can be much larger, up to around 5\%.
Such a large effect needs to be taken into account in precise studies
of \bbg events where the experimental errors can be comparable with, or
smaller than, this size.

For example, tests of the flavour-independence
of strong interactions involve measurements of the ratios $r^Q(X)$ =
$X^Q/X^{uds}$
of a 3-jet observable $X$ in \QQg versus $q_l\overline{q_l}g$ events.
Currently the experimental errors on $r^b(X)$ are of the order of 1-2\%,
and $b$-mass
effects are clearly visible in the data~\cite{phil,delphi}.
By contrast, the errors on $r^c(X)$ are much larger than 1\% and any
$c$-mass effects
are not discernible.
Hence, in recent measurements the NLO massive calculations have
been employed to correct $r^b(X)$  for the $b$-mass effects, so as to
determine the ratio
of strong couplings, $\alp^b/\alp^{uds}$~\cite{delphi,sld,opal}, and test the
{\it ansatz} of flavour-independence of strong interactions.

An alternative, and \apriori equally valid, approach is to assume that
strong interactions are flavour independent,
and use the sensitivity of event-shape observables
to mass effects to determine the $b$-mass itself.
In the theoretical prediction one has the freedom to choose the
renormalization scheme which defines the quark mass. For example,
one can write the NLO result in terms of either the perturbative
pole mass $M_b$ or the `running' mass $m_b(\mu)$.
The latter is defined by the modified minimal subtraction
(\msbar) scheme~\cite{msbar} employed to renormalize the mass at a scale
$\mu$.
At the  \z0 scale, $M_Z$, 
the running mass is preferable because large logarithms 
of the form $\ln(M_b^2/M_Z^2)$ are absorbed in $m_b(M_Z)$, 
and the perturbative expansion is thus improved.
The DELPHI Collaboration has recently studied the 3-jet-rate
$R_3(y_c)$, where $R_3$ was determined using the
Durham (D) jet-finding algorithm~\cite{durham}, and $y_c$ is the
scaled-invariant-mass
criterion which determines the jet multiplicity. From their measurement
of
$r^b(R_3)$ at $y_c$ = 0.02 DELPHI obtained~\cite{delphi}:
\begin{equation}
m_b(M_Z)= 2.67 \pm 0.25 ({\rm exp.})\pm0.34({\rm frag.})
\pm0.27 ({\rm theor.})\;{\rm GeV}/c^2.
\label{eq:delphi}
\end{equation}

Such 3-jet-event observables have been used for many years
to determine $\alpha_s(M_Z)$ from inclusive-flavour \ep
annihilation events~\cite{alphas}.
Though the \alp value obtained by fitting a NLO pQCD
calculation to any one measured observable can have quite a small
experimental
error, $\leq0.001$ for some observables, there is a strong dependence
of the
fitted \alpmz value on the choice of observable~\cite{sldalp,lepalp}.
This spread of values leads to a large and dominant uncertainty on
\alpmz determined
with this technique~\cite{alphas}. Since non-perturbative
effects are supposedly taken into account in these measurements, usually by
applying corrections
based on well-tested hadronisation models~\cite{models}, a consistent
description within the framework of pQCD is viable only if one
postulates large
(and uncalculated) next-to-next-to-leading-order (NNLO) contributions to
the
observables. Hence, in this picture, the spread in \alpmz values
determined at NLO
results from the omission of the uncalculated higher-order terms.
Furthermore, it can be
argued that a strong dependence of a NLO calculation on the
renormalisation scale
is generally a sign of large NNLO contributions. Such a dependence is
indeed
observed for most of the observables~\cite{sldalp,lepalp,optscale}, and
supports the
previous interpretation, though there is little consensus on a procedure
for quantifying
the scale-dependence of measurements of \alpmz.

The DELPHI determination of \mbz\ (Eq.~\ref{eq:delphi}) is based on a 
ratio of 3-jet-event observables calculated at NLO.
Given this apparently very precise result derived from one observable,
it is interesting to consider the possible effect of NNLO contributions.
Naively one might expect any potentially sizeable
effects in the numerator and denominator largely to cancel. However, a
residual
uncertainty at only the 2\% level on $r^b$ corresponds (Section 3) to a
0.5 GeV/$c^2$
error on \mbz, which is comparable with the quoted total error on the
DELPHI
measurement. For the purpose of investigation we have
studied the extraction of \mbz from the ratios $r^b(R_3)$, where 
$R_3$ was determined using six different infra-red- and
collinear-safe jet-finding algorithms. As in the case of \alpmz
measurements
using such observables, the study of an ensemble of results from
different observables,
all calculated at NLO, may uncover systematic effects relating to
the uncalculated NNLO contributions.
 We used the Durham and Geneva (G)~\cite{bkss}
schemes, and the E, E0, P and P0 variations of the JADE
algorithm~\cite{jade}
to evaluate the $b$-mass-dependent NLO pQCD predictions,
and compared them with
the corresponding experimental measurements published by the SLD
Collaboration~\cite{sld}.

In Section 2 we outline the theoretical framework and briefly describe
the NLO
calculations used here. In Section 3 we compare the calculations with
the data and extract values of \mbz using each jet algorithm in turn.
We compare the results obtained using the different jet algorithms, and
discuss the systematic uncertainties. 
In Section 4 we combine these results to obtain our best estimate
of the central \mbz value and error
by taking correlations into account. These results supercede the
preliminary results presented in~\cite{phil}. 

\vskip 1truecm

\section{Theoretical Framework}

\vskip .5truecm

\noindent
In this section we describe the computation of the double ratio:
\begin{equation}
r^b(y_c)\quad\equiv\quad R_3^b(y_c)/R_3^{uds}(y_c),
\label{eq:observable}
\end{equation}
where we define $R_3^q(y_c)$ to be the fraction of events classified as
containing
3 {\it or more} jets with a particular jet-finding algorithm at a particular
$y_c$ value.
The event flavour is defined by the flavour of the primary quarks that
couple to the
\z0. This definition means that events of the type
$Z^0\to q_l\bar{q_l}g\to q_l\bar{q_l}b\bar{b}$ are classified
as light-quark events
{\footnote{This distinction is possible only because
the contribution to $R_3^q$ from the interference
of the amplitudes for $Z^0\to q_l\bar{q_l}g\to q_l\bar{q_l}b\bar{b}$
and $Z^0\to b\bar{b}g^\ast\to b\bar{b}q_l\bar{q_l}$ is negligible.
In fact the contribution cancels exactly in $R_3^b$ if one neglects the 
masses of the $u,d,s,c$ quarks and the sum over the light quarks $q_l$
includes only full isospin doublets.}}
$R_3^b(y_c)$ and $R_3^{uds}(y_c)$ can be written to
NLO accuracy:
\bea
R_3^q(y_c)=\frac{\alpha_s}{2\pi}A^q(y_c)+
\left(\frac{\alpha_s}{2\pi}\right)^2 (B^q(y_c)+C^q(y_c)) +
O\left(\alpha_s^3\right),
\eea
where the coefficients $A$ ($B$) represent the LO (NLO) contribution to
3-jet production, and the coefficients $C$ represent the LO contribution
to 4-jet production.
Thus we have (we suppress here the argument $y_c$):
\def\Auds{A^{uds}}
\def\Ab{A^b}
\bea
\label{eq:rbexpand}
r^b = \frac{\Ab}{\Auds}+\frac{\alpha_s}{2\pi}
\left(\frac{B^b+C^b}{\Auds}-\frac{B^{uds}+C^{uds}}{\Auds}
\frac{\Ab}{\Auds}\right)+O\left(\alpha_s^2\right).
\eea

Some comments are in order about the generalization of observables 
originally defined for massless quarks to account for quark masses. Because 
of the fact that, for massless quarks, energy and the modulus of the 
three-momentum can be interchanged, the generalization to massive quarks 
contains a certain degree of freedom. For massive quarks 
one should use a definition which is infrared safe and does not
distinguish between a hard parton and the hard parton substituted by 
two collinear ones, or a hard parton and soft parton. 
(See for example~\cite{Gottschalk} where this is discussed for the example of 
the thrust.) 
While this is not a problem for the Durham, Geneva, and E algorithms one should
be careful when using algorithms which use a recombination procedure that
keeps the recombined four-momentum massless, in particular the E0, P, and
P0 algorithms. Replacing a massive quark by a hard parton and a soft one,
or two collinear partons, would yield a massless four-momentum vector after
the recombination, instead of getting the original four-momentum back. 
These algorithms hence `feel' soft and collinear partons if they
are naively applied for massive quarks. (Fortunately this
problem does not occur if further splitting is considered.)

One possibility would be to modify these algorithms in some fashion
so as to keep the hadronisation corrections small; this was in fact
the original motivation for developing the Durham and Geneva algorithms. 
Here we preserve the definition of the algorithms so as to be able to 
compare with the published results of the SLD analysis.
The SLD measurements are corrected to the parton-level, where 
the naive jet definition is applied even if one of
the four-momenta is massive.
We are hence able to extract consistently the $b$ mass from these measurements
obtained with the E0,P, and P0 jet algorithms by treating the $b$ quark 
as massive in our theoretical prediction.

\subsection{3-jet Contributions}

\noindent
For 3-jet production
the LO massless ($\Auds$) and massive ($\Ab$) \cite{Io78} 
as well as the NLO massless ($B^{uds}$) \cite{nlo} coefficients are well
known.  
In order to calculate the massive NLO coefficients $B^b$
we need the matrix elements for
\begin{equation}
  \label{eq:3-partons}
  e^+e^-\to(\gamma^\ast,Z^\ast)\to b\bar b g
\end{equation}
to order $\as^2$, as well as the matrix elements for the parton
processes
\begin{equation}
  \label{eq:4-partons}
  e^+e^-\to(\gamma^\ast,Z^\ast)\to b\bar bgg,
  \, b\bar b q_l\bar{q_l},\, bb\bar b\bar b.
\end{equation}
In the calculation of the virtual corrections to
Eq.~\ref{eq:3-partons}
both ultra-violet (UV) and infra-red (IR) singularities are encountered.
The
UV singularities are removed by the usual renormalization procedure of
the
mass parameter and the QCD coupling $\as$.

The IR singularities are cancelled by
the real contributions from the processes listed in \eq{eq:4-partons}.
It is worthwhile adding some remarks about this cancellation.
Today it is more or less standard to regulate the IR
divergences in the framework of dimensional regularization. To cancel
the divergences in the virtual corrections one must then integrate the
real contributions over some regions of phase-space in $d$ dimensions.
More precisely one must integrate over the regions where a gluon is
soft or two massless partons are collinear.
In general this would be a formidable task.
Therefore several techniques
have been developed in the past (see for example
\cite{GiGl92,CaSe96b,CaSe97,GiGlKo93})
to simplify this problem using the general
factorization properties of QCD amplitudes. In the calculation reported
in \cite{arnd} on which the current paper is based the so called
phase-space-slicing
method \cite{GiGl92} was used.
The same is true for the results presented in
\cite{rodrigo} on which the DELPHI analysis~\cite{delphi} is based. 
In the calculation given in \cite{nason} an alternative,
the so-called subtraction method, was used. 

In the simplest version of the
phase-space slicing method one separates the `soft' and `collinear'
regions
(often called `unresolved regions') from the rest of the phase-space
(`resolved regions') by demanding a minimal invariant mass-squared
$\smin$ for all pairs of partons.
In the soft and collinear regions the squared matrix elements can be
approximated by the use of the soft and collinear factorization which is 
valid in the appropriate limits. After this
simplification the relevant part of the squared matrix elements can be
integrated analytically in $d$ dimensions.
The phase-space integration over the resolved regions
can be done numerically in four dimensions. In the case of
massive quarks the phase-space slicing method must be modified, although
the 
basic features are the same. In particular, the `slicing' between
the soft/collinear regions  and the regions where all partons are hard
can still be parametrized in terms of one variable $\smin$.

The approximation used in the unresolved region is only valid
for small values of $\smin$. On the other hand for small $\smin$ large
cancellations between the numerically integrated and the analytically
integrated parts will arise leading to possible errors in the sum of the
two.
Note that the artificial cut parametrized by $\smin$ is not related to
any physical cut. Thus the theoretical prediction must
be independent of $\smin$. In practice, for the NLO coefficient
$B^b$ this will be true up to corrections of the order of $\smin/s$.
With the value $\smin =0.5 {\mbox{ GeV}}^2$, which we have
used in our calculation, the systematic
error in $B^b$ due to the phase-space slicing method is negligible
compared with the numerical error due to the numerical integration, which is
itself negligibly small.

Note that although free of collinear singularities 
in the case of massive 
quarks, one must also include the contribution of the reaction 
$e^+e^-\to(\gamma^\ast,Z^\ast)\to bb\bar b\bar b$ to the 3-jet 
rate. 
The collinear singularity appearing for massless quarks shows up as 
$\ln(m_b)$ for massive quarks.
Only if this contribution is included is the 3-jet contribution free
of collinear singularities in the limit of vanishing $b$-quark mass. 
The logarithm cancels against the logarithm coming from the virtual
corrections to the gluon propagator due to a $b$-quark loop.
Needless to say, this reaction contributes also to the 4-jet rate.

Note also that in the reaction $e^+e^-\to(\gamma^\ast,Z^\ast)\to b\bar bgg$
collinear singularities (due to collinear configurations of a 
(anti-)quark gluon pair) that appear for massless quarks give rise only to
logarithms of the mass. This explains why it is difficult to calculate 
$R_3^b$ numerically for small values of $m_b$ without further analytical 
work. In the virtual correction these logarithms are explicit. In the real
contribution they must be obtained numerically 
from the phase space integration, which
becomes more difficult for smaller masses due to numerical instabilities.

For technical reasons it is easier to perform the calculation of $r^b$
first in the pole mass scheme, and switch to the running mass
afterwards.
The relation between the pole mass $M_b$ and the \msbar\  mass 
$m_b(\mu)$ reads
\bea
M_b=m_b(\mu)+\delta m_b(\mu),
\eea
where, to order $\as$,
\bea
\delta m_b(\mu) = \frac{\alpha_s}{\pi}\left(\frac{4}{3}-
\ln\frac{m_b^2(\mu)}{\mu^2}\right)m_b(\mu).
\eea
This implies the following relation between $r^b$ in both mass
renormalization schemes~\cite{spira}:
\bea
\label{eq:polrun}
r^b(M_b) = r^b(m_b(\mu)) + \frac{1}{\Auds}
\delta m_b(\mu) \frac{d\Ab(m_b(\mu))}{dm_b(\mu)} +O(\as^2).
\eea
The mass dependence of $A^b$ can be written as 
$A^b(m_b) =A^{uds} + \delta A^b(m_b)m_b^2/s$, which defines
the function  $\delta A^b(m_b)$.
For $m_b^2 \ll s$,  $\delta A^b$ depends only weakly 
on $m_b$. Ignoring this residual mass dependence we have 
\bea
\label{eq:approx}
\frac{1}{\Auds}
\delta m_b(\mu) \frac{d\Ab(m_m(\mu))}{dm_b(\mu)}
\approx
{1\over\Auds }
\left[\Ab\left(\sqrt{m_b^2(\mu)+2m_b(\mu)\delta m_b(\mu)}\right)
-\Ab(m_b(\mu))\right],
\eea
and we use the r.h.s. of (\ref{eq:approx}) to convert the results
for $r^b$ from the pole mass to the \msbar\ mass renormalization scheme.
This excellent approximation avoids the calculation of
the derivative $d\Ab/dm_b$ for each algorithm.

\subsection{4-jet Contributions}

Both the massless \cite{nlo} and massive \cite{BMM92}  
LO 4-jet fraction contributions ($C$) are well known. 
Recently, the massless 4-jet fraction has been computed
to NLO \cite{DiSi97}. These corrections, which are of order
$\alpha_s^3$ and therefore not included in our prediction
for $r^b$, can change, depending on the jet algorithm,
the values of the massless $C$ coefficients by up to 100\%. 
Note that part of these large NNLO corrections to $r^b$ 
will cancel between the massless and (yet unknown) 
massive $O\left(\alpha_s^3\right)$ $C$ 
coefficients entering (\ref{eq:rbexpand}).

\subsection{Calculation of $r^b$}

\begin{figure}[ht]
 \hspace*{0.5cm}
   \epsfxsize=5.625in
   \epsfysize=7.5in
\begin{center}\mbox{\epsffile{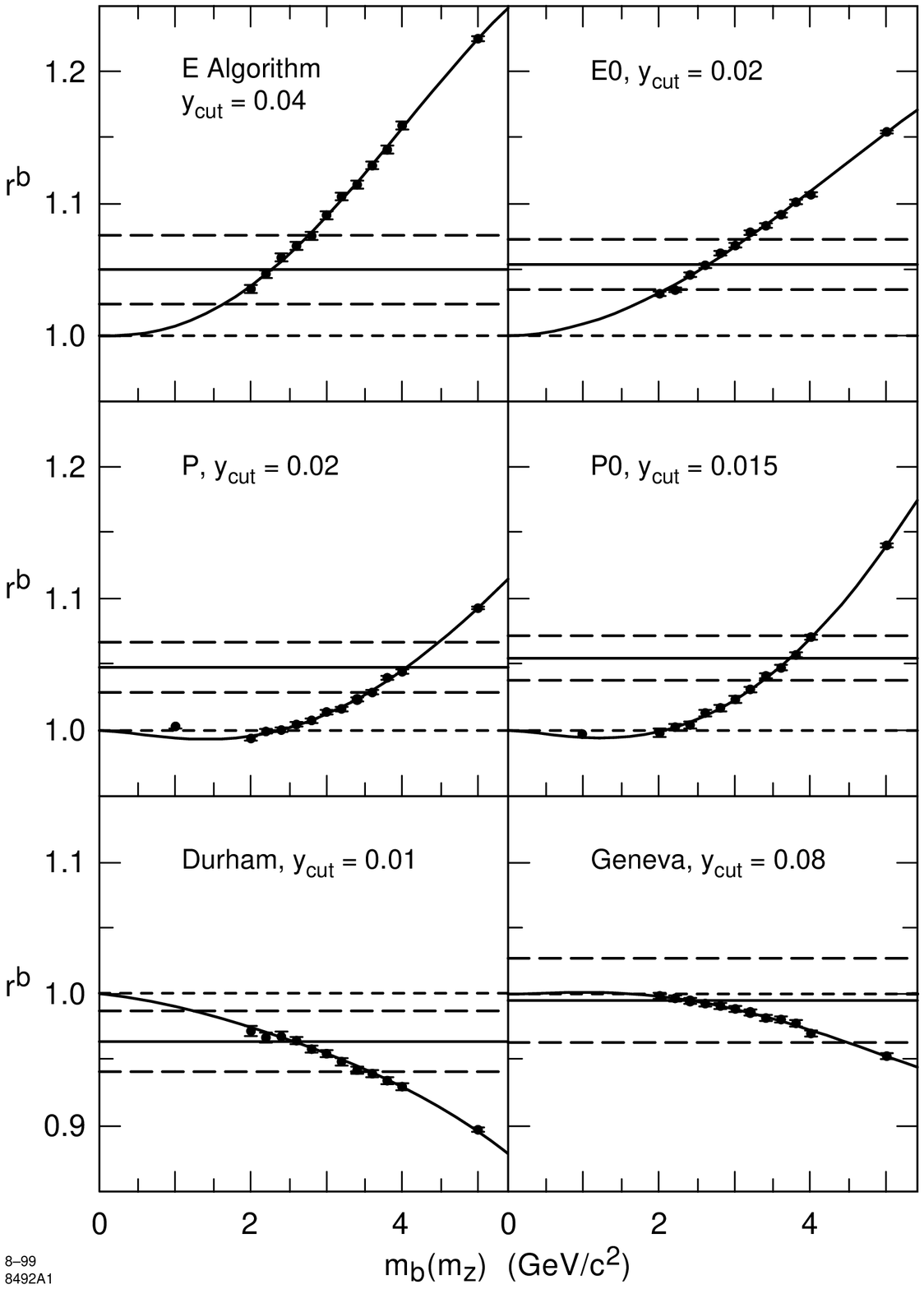}}\end{center}
\vspace*{-0.75cm}
  \caption{
 \label{sixfigs}
The $R_3^b/R_3^{uds}$ ratios measured by SLD for each of the six jet
finding
algorithms (horizontal bands) compared with the predicted dependence on
the
running $b$-quark mass, $m_b(M_Z)$.
    }
\end{figure}

The calculation of $r^b$ was performed,
for each of the six jet algorithms at the optimal $y_c$ values
(discussed below),
for \mbz values in the range
$2.0 \leq \mbz \leq 5.0$ GeV/$c^2$.
These predictions are shown as points in Fig.~\ref{sixfigs}.
For the P and P0 algorithms
we also calculated a point at $m_b(M_Z)$ = 1 GeV/$c^2$ so as to
constrain better the extrapolation $r^b$ $\rightarrow$ 1 as $m_b(M_Z)$
$\rightarrow$ 0.
The renormalization scale $\mu$ was set to $\sqrt{s}$ and
we used $\alpha_s(M_Z)=0.118$.
The dependence of $r^b$ on the renormalization
scale $\mu$ is trivial if $r^b$ is expressed in terms of the
pole mass $M_b$; it enters only through
the running of $\alpha_s$. An additional $\mu$ dependence
is introduced if one switches to the running mass $m_b(\mu)$
by using Eq. (\ref{eq:polrun}). The theoretical uncertainty on
$m_b(M_Z)$ due to the choice of the 
renormalization scale will be discussed in section 3.

The function
\begin{equation}
\label{eq:fit}
f(m)\quad=\quad
1+\alpha \frac{m^2}{s}
 + \beta \frac{m^2}{s} \ln \left(\frac{m^2}{s}\right) + 
\gamma \frac{m^4}{s^2}
\end{equation}
where $\alpha$, $\beta$
and $\gamma$ are free parameters, was fitted to these points.
The {\it ansatz} (\ref{eq:fit}) can be jusified as follows:
1) As $m\to 0$, the massive fraction $R^b_3$ approaches the massless
one{\footnote{This is true up to differences induced by 
triangle diagrams \cite{HaKuYa91}, which lead to deviations of the
$r^b(m=0)$ from 1 of less than 0.1\%.}}
$R_3^{uds}$; 
2) since $m^2\ll s$, it is a very good approximation to keep 
only the leading terms in $m^2/s$.
The fitted parameter values are listed in Table \ref{tfitpar}; the
functions are shown in Fig. \ref{sixfigs} and provide a
good description of the mass dependence of the calculations.

\begin{table}[ht]
\begin{center}
\begin{tabular}{|c|c|ccc|}
\hline
Algorithm  & $y_c$ & $\alpha$ & $\beta$  & $\gamma$ \\
\hline
  E   &  0.040 & 207.6  &   16.10  &    $-13029.9$  \\
  E0  &  0.020 & 42.2  &   $-3.58$  &    $-3881.3$   \\
  P   &  0.020 & 194.6  &   26.17 &    $-3855.9$   \\
  P0  &  0.015 & 206.0  &   26.73  &    $-1317.5$   \\
  D   &  0.010 & 79.3  &   17.16  &    $-4610.8$   \\
  G   &  0.080 & $-89.6$  &   $-11.04$ &     3229.9   \\
\hline
\end{tabular}
\caption{Fitted parameters of the function
$f(m)=1+\alpha m^2/s + \beta (m^2/s) \ln (m^2/s) + \gamma m^4/s^2$ 
for each jet algorithm.
}
\label{tfitpar}
\end{center}
\end{table}

It can be seen that the \mbz-dependence varies according to the jet
algorithm.
For \mbz $\geq$ 2.0 GeV/$c^2$, $r^b>1$ and the slope is
positive for the E, E0, P and P0 cases, whereas $r^b<1$ and the
slope is negative for the D and G cases.
This can be
understood qualitatively in terms of two competing physical origins.
First, the non-zero $b$-mass tends to cause a phase-space suppression of
gluon
emission relative to the massless quark case, implying $r^b$ $<$ 1. 
Second, for a given kinematic configuration,
the large $b$-mass tends to enhance the invariant mass of a local
quark-gluon
pair relative to the massless quark case.
Since the JADE family of jet
algorithms is based on a clustering metric that is closely related to
invariant mass, for fixed $y_c$ the two partons are more likely to be
resolved
as separate jets when the quark is massive, implying $r^b \geq 1$.
By contrast, the
clustering metric used in the Durham and Geneva algorithms is less
sensitive
to this kinematic effect, the phase-space suppression dominates, and
$r^b \leq 1$. For increasing values of $y_c$ one expects both effects
to diminish in importance and $r^b$ \ra 1.
For the D algorithm this has been observed in the DELPHI
study~\cite{delphi}.

\vskip 1truecm

\section{Extraction of the $b$-Quark Mass}

\vskip .5truecm

\noindent
We used measurements of $r^b$ published~\cite{sld} by the SLD
Collaboration.
These measurements are based on a sample of 150,000 hadronic \z0 decays
recorded between 1993 and 1995, for which the original 120-million pixel
CCD vertex detector
was used for event flavour separation. SLD has subsequently recorded a
further 400,000
\z0 decays with a new 307-million-pixel vertex detector, and it would be
straightforward to repeat the present analysis when the new data are
made available.
Though not as statistically powerful as the DELPHI result for the
D jet algorithm, the SLD published data include results for the six
different
jet algorithms D, G, E, E0, P and P0, and are hence suitable for this
study
of possible observable-dependent systematic effects.

Full details of the experimental procedure are given in~\cite{sld}.
Briefly, \ep $\rightarrow$
hadrons events were selected, and a flavour-tagging algorithm was
applied to select
samples of events of primary $b$, $c$, and $uds$ quark flavour.
The algorithm was based on the mass and momentum
of secondary decay vertices reconstructed using the vertex detector.
Light-quark ($uds$) events rarely contain reconstructed
secondary decay vertices, and these typically result from
strange particle decays and are of low mass.
Conversely, \bb events typically contain high-mass vertices from
$B$-hadron decays.
The purity of the $b$-tagged ($uds$-tagged) event sample was 90\% (91\%)
respectively.

Each jet-finding algorithm was applied in turn to the $uds$- and
$b$-tagged samples
and, for each algorithm, the ratios (Eq.~\ref{eq:observable}) were
formed.
The ratio is an attractive quantity as many of the experimental
and theoretical systematic uncertainties effectively cancel.
Each ratio was then corrected for the effects of
detector acceptance and resolution, the bias of the flavour tag to
select
preferentially 2-jet rather than 3-jet events, 
the flavour compositions, and
hadronisation effects.
For each algorithm an `optimal' $y_c$ value was selected so as to
minimise the
combined statistical and experimental systematic error.

The measured $r^b$ values and the associated errors are listed in
Table~\ref{tab:sldr}~\cite{sld}. The central values and statistical
errors are also shown in Fig.~\ref{sixfigs}.
The set of $r^b$ values is not consistent with unity, which
indicates that the $b$-mass effects are significant.
Furthermore, a systematic algorithmic dependence is apparent,
with $r^b \geq 1$ for
the JADE family of algorithms and $r^b\leq 1$ for the D and G
algorithms, in agreement with the expectations discussed in Section 2.

\begin{table}[ht]
\begin{center}
\begin{tabular}{|c|c|c|c|c|c|}
\hline
Algorithm  & $y_c$ & $r^b$  & stat. & exp. syst. & had.  \\
\hline
  E   & 0.040 & 1.050 & 0.026 & $^{+0.038}_{-0.042}$ &
$^{+0.011}_{-0.046}$ \\
  E0  & 0.020 & 1.054 & 0.019 & $^{+0.030}_{-0.037}$ &
$^{+0.007}_{-0.045}$ \\
  P   & 0.020 & 1.048 & 0.019 & $^{+0.027}_{-0.037}$ &
$^{+0.002}_{-0.026}$ \\
  P0  & 0.015 & 1.055 & 0.017 & $^{+0.028}_{-0.035}$ &
$^{+0.007}_{-0.037}$ \\
  D   & 0.010 & 0.964 & 0.023 & $^{+0.038}_{-0.041}$ &
$^{+0.001}_{-0.006}$ \\
  G   & 0.080 & 0.995 & 0.032 & $^{+0.035}_{-0.036}$ &
$^{+0.020}_{-0.008}$ \\
\hline
\end{tabular}
\caption{
SLD measured values and errors of $r^b$.
}
\label{tab:sldr}
\end{center}
\end{table}

For each jet algorithm, by comparing the theoretical curve in
Fig.~\ref{sixfigs}
with the SLD data, one can read off the preferred \mbz value.
The central values are listed in Table~\ref{tbmass}. In each case
upper and lower statistical errors were evaluated from the crossing
points of
the error band with the theoretical prediction,
except in the case of the G
algorithm, for which the upper statistical bound is consistent with
$m_b=0$; in
this case an error equal to the central value was assigned.
Each experimental systematic error on $r^b$~\cite{sld} was similarly
transformed into a
systematic error on \mbz and the sum in quadrature is listed in
Table~\ref{tbmass}.
Hadronisation uncertainties~\cite{sld} were evaluated in a similar
fashion and
are listed in Table~\ref{tbmass}.

Additional theoretical uncertainties were investigated by varying the
value of $\alpha_s(M_Z)$ within the range $0.115\leq$\alpmz$\leq0.121$.
The corresponding changes in \mbz were at the level of 
$\pm(10-20)$ MeV/$c^2$. The renormalisation scale was also varied 
within the range $M_Z/2\leq\mu\leq 2M_Z$. The corresponding changes in 
\mbz were at the level of $\pm200$MeV/$c^2$ or less. For each
algorithm these uncertainties were added in quadrature to define a
theoretical uncertainty, which is listed in Table~\ref{tbmass}.

\begin{table}
\begin{center}
\begin{tabular}{|c|rrrrr|}
\hline
Algorithm  & $m_b(M_Z)$ & stat. & exp. syst. &  had. & theor. \\
\hline
   E     &   2.271 & $+$0.488   & $+$0.734   & $+$0.217 & $+0.194$\\
         &         & $-$0.629   & $-$0.952   & $-$1.483 & $-0.189$\\
\hline
   E0    &   2.642 & $+$0.493   & $+$0.789   & $+$0.187 & $+0.213$\\
         &         & $-$0.562   & $-$1.082   & $-$1.637 & $-0.226$\\
\hline
   P     &   4.067 & $+$0.423   & $+$0.616   & $+$0.048 & $+0.047$\\
         &         & $-$0.504   & $-$0.981   & $-$0.723 & $-0.021$\\
\hline
   P0    &   3.728 & $+$0.307   & $+$0.515   & $+$0.132 & $+0.056$\\
         &         & $-$0.361   & $-$0.738   & $-$0.911 & $-0.043$\\
\hline
 D       &   2.509 & $+$1.028   & $+$1.879   & $+$0.287 & $+0.170$\\
         &         & $-$1.255   & $-$2.001   & $-$0.049 & $-0.195$\\
\hline
 G       &   2.415 & $+$2.075   & $+$2.761   & $+$0.691 & $+0.195$\\
         &         & $-$2.415   & $-$2.415   & $-$2.415 & $-0.078$\\
\hline
\end{tabular}
\caption{Values of the running $b$-quark mass extracted from the SLD
measurement of $r^b$ for each of the six jet algorithms.
}
\label{tbmass}
\end{center}
\end{table}

The six measured $b$-quark masses range from 2.3 to 4.1 GeV/c$^2$, with
an r.m.s.
deviation of 0.7 GeV/c$^2$;  this scatter is larger than one might
expect from
these data given the strong correlations between measurements using
different jet algorithms, suggesting some additional source of
uncertainty.
In order to quantify this issue we evaluated
the statistical correlations among the $r^b$ values determined using
different jet
algorithms. We repeated the analysis on subsets of both the data
and the simulated data and calculated the correlation coefficients
empirically.
The data and simulation gave consistent results, and the average
correlation
coefficients are listed in Table \ref{tcorrels}.
Each has a statistical uncertainty of $\pm$0.03.
The four JADE-like algorithms show strong correlations with each other,
in the range 0.65--0.84, as might be expected. Correlations between
other pairs of
algorithms are weaker, in the range 0.41--0.65.

\begin{table}
\begin{center}
\begin{tabular}{|c|rrrrrr|}
\hline
Algorithm & E  &  E0  &  P   &  P0  &   D  &  G  \\
\hline
  E     & 1.00 & 0.70 & 0.67 & 0.65 & 0.61 & 0.49  \\
  E0    &      & 1.00 & 0.84 & 0.82 & 0.61 & 0.49  \\
  P     &      &      & 1.00 & 0.71 & 0.65 & 0.56  \\
  P0    &      &      &      & 1.00 & 0.52 & 0.41  \\
  D     &      &      &      &      & 1.00 & 0.64  \\
  G     &      &      &      &      &      & 1.00  \\
\hline
\end{tabular}
\caption{Statistical correlation coefficients between $r^b$
measurements for each pair of jet-finding algorithms.}
\label{tcorrels}
\end{center}
\end{table}

We evaluated
\begin{equation}
\chi^2 = \Sigma_{ij} (r^b_i - f_i(\mbz)) (V^{-1})_{ij} (r^b_j-f_j(\mbz)),
\end{equation}
where $r^b_i$ ($f_i$) are the measured (calculated) double
ratios,
$i,j=$ E, E0, P, P0, D, G, the error matrix is defined by
$V_{ij} = c_{ij}\sigma_i \sigma_j$, $c_{ij}$ is the correlation coefficient
given in Table \ref{tcorrels}, and $\sigma_i$ is the quadratic sum of the
data and Monte Carlo statistical errors on $r^b_i$.
For values of \mbz around 2.9 GeV/$c^2$, which is the average of
the results shown in Table~\ref{tbmass}, we obtained $\chi^2$ $\simeq$ 
38/6, which indicates an inconsistency among the results from the 
different algorithms.
We minimised $\chi^2$ with respect to variation of \mbz and
obtained $\chi^2$ = 26/5 for \mbz = $0.9\pm0.7$ (stat.) GeV/$c^2$, which is still 
unacceptably high.
The best-fit $\chi^2$ value is insensitive to variations of the $c_{ij}$
within
their uncertainties, and to (simultaneous) systematic shifts of the
measured
$r_i$ within the experimental systematic errors and hadronisation and
theoretical uncertainties.
The experimental systematic errors, which are dominated by uncertainties in the
flavour composition of the samples, and the hadronisation uncertainties
were assumed to be 100\% correlated among all algorithms and were omitted from 
the $\chi^2$ calculation; theoretical uncertainties were also omitted.

We repeated this minimisation procedure and omitted in turn the 
measurement based on each of the six algorithms. In no case did we
obtain a $\chi^2$ value better than 12, which corresponds to a
confidence level of 1.7\%. We then omitted pairs of
measurements in turn. Fits with $\chi^2$ $<$ 9, \ie $>$6\%
confidence level, were obtained only for two of these 15 cases in which
both the E and E0, or both the P and P0, 
algorithms were omitted; the corresponding mass
values were 2.5 and 3.5 GeV/$c^2$, respectively.
To the extent that the hadronisation and
theoretical uncertainties have been properly estimated, we do not have
\apriori justification for omitting any particular algorithm(s).
We do note, however, that algorithms in
the JADE family 
have a significantly worse soft gluon behaviour 
than the D and G algorithms \cite{bkss}. 
The former algorithms tend to combine soft gluons to  
form an `artificial' jet at values of $y_c$ that
are not small, which may cause large higher-order
perturbative corrections even for moderate
values of $y_c$.

The $\chi^2$ value is, however, quite sensitive to small changes in the
measured $r^b_i$ or predicted $f_i$.
As an exercise, we have postulated an additional uncertainty
of size $\epsilon$ which is uncorrelated between different jet
algorithms.
Under the hypothesis $m_b=2.94$ GeV/c$^2$, which is the average of the values
listed in
Table~\ref{tbmass}, for $\epsilon=0.015$ the $\chi^2$  value
is 10.5; for $\epsilon=0.02$ the $\chi^2$  value is 7.1, which is
acceptable.
For $\epsilon \geq 0.02$ a fit for \mbz yields a stable value of
roughly 2.6 GeV/c$^2$, indicating that
a consistent \mbz value {\it can} be obtained provided that there exist
{\it additional} uncertainties, uncorrelated between jet algorithms, at
the level of
2\% on $r^b$.
A 2\% error on the predicted $r^b$ corresponds to an error of $\sim$0.5
GeV/c$^2$
on the extracted value of \mbz from a given algorithm, and would
roughly account for the 0.7 GeV/c$^2$ r.m.s. deviation
among the values in Table~\ref{tbmass}.

We suspect that the most likely
source of the inconsistency among results for the different jet
algorithms
is the missing higher-order perturbative contributions to $r^b$.
As we have shown, these would have to be only at the level
of 2\% in order to resolve the inconsistency.
Possible NNLO contributions to $r^b$
of this small magnitude are not \apriori unexpected;
5--10\% level NNLO contributions are implied by the scatter among
\alpmz values
determined using these and closely-related event-shape
observables~\cite{alphas} which
form the numerator and denominator of $r^b$.

\vskip 1truecm

\section{Summary and Conclusions}

\vskip .5truecm

\noindent
We have studied the determination of the running $b$-quark mass by comparing
NLO perturbative QCD calculations of the $\geq3$-jet ratio $r^b$ = 
$R_3^b/R_3^{uds}$ with data from the SLD Collaboration. We used six different
infra-red- and collinear-safe jet-finding algorithms in order to study
systematic effects.
We find algorithm-dependent values of \mbz in the range 
$2.3<\mbz<4.1$ GeV/$c^2$.
The value determined using the Durham algorithm is consistent with that
reported in~\cite{delphi}. 

We quantified the statistical, experimental
systematic, hadronisation and additional theoretical uncertainties,
and attempted to obtain a best-fit \mbz value by minimising $\chi^2$, taking
statistical correlations between the results from the different algorithms
into account. 
We could not obtain an acceptable best-fit $\chi^2$ value unless, in two cases,
we omitted two algorithms from consideration. In the absence of
an \apriori reason to do this we retained all six algorithms in order
to investigate possible additional systematic effects.
We were able to obtain an acceptable value of $\chi^2$, and a stable
value \mbz $\simeq$ 2.6 GeV/$c^2$, provided that
we postulated (an) additional source(s) of uncertainty 
of relative size $\geq2$\% on $r^b$, which is uncorrelated
between algorithms. 
We are unable to account for the origin of such an uncertainty, but
speculate that it may be due to uncalculated higher-order pQCD contributions.

We now discuss the assignment of a single value of \mbz. Taking an
unweighted average of the $m_b$ values in Table~\ref{tbmass}, yields
$m_b(M_Z)$ = 2.94$^{+0.80}_{-0.95}$ (stat.) $^{+1.22}_{-1.36}$ (syst.)
$^{+0.26}_{-1.20}$ (had.)  $^{+0.11}_{-0.12}$ (theor.)
$\pm0.69$ (r.m.s.) GeV/$c^2$,
where we include the r.m.s. deviation as an additional error.
Though well defined, this procedure does
not make full use of the information contained in the six measurements.
The $\chi^2$ minimisation procedure does take into account the full
statistical covariance matrix, as well as correlations in the
systematic error and hadronisation uncertainties. 
Since the resulting $\chi^2$ is acceptable, and the fitted \mbz value 
is stable for any additional uncorrelated uncertainties of size 
$\epsilon$ $\geq$ 0.02, we choose $\epsilon=0.02$, and obtain 
$\chi^2$ = 6.5/5 with
\begin{equation}
m_b(M_Z) = 2.56 \pm 0.27 {\rm (stat.)} ^{+0.28}_{-0.38} {\rm (syst.)}
                ^{+0.49}_{-1.48}  (\rm theor.) \; {\rm GeV}/c^2.
\end{equation}
We consider that this represents our best estimate of the running 
$b$-quark mass using the SLD data. 

The statistical error on \mbz is substantially reduced by the
correlations among
the six individual $r^b$ results. The experimental systematic error is
also
reduced by the fact that a given shift of $r^b$ causes the \mbz values
for the E, E0, P and P0 algorithms to shift
in opposite directions to those for the D and G algorithms.
The theoretical uncertainty comprises the sum in quadrature of the 
hadronisation uncertainty ($^{+0.25}_{-1.42}$ GeV/$c^2$), 
and the propagation of the uncorrelated error of $\pm0.02$
on each $r^b$ ($\pm 0.42$ GeV/$c^2$).
Variations of \alp and $\mu$
contribute an uncertainty of $^{+0.14}_{-0.12}$ GeV/$c^2$;
under the assumption that the `$\epsilon$ error' results from 
uncalculated higher-order
pQCD contributions the effects represented by these
variations are already `counted', and   
we have not added them in quadrature with the other theoretical
uncertainties. Their inclusion does not change the central \mbz 
value and increases the total theoretical uncertainty to
$^{+0.51}_{-1.49}$ GeV/$c^2$.

Our result is in agreement with that from~\cite{delphi}. The latter
measurement has a significantly smaller theoretical uncertainty.
For the Durham algorithm alone we would obtain an uncertainty of
similar size, but our study of six different jet algorithms has
revealed additional systematic effects which warrant further
investigation. 

\vskip 1truecm

\noindent
\section*{Acknowledgements}

\vskip .5truecm

\noindent
P.N.B., D.M. and N.O. thank colleagues 
in the SLD Collaboration for their support
for this work.
We thank Werner Bernreuther, Otmar Biebel, and 
Mike Seymour for helpful discussions.

\vfill\eject

\end{document}